\documentclass[aps,reprint,superscriptaddress, nobalancelastpage, raggedbottom]{revtex4-2}

 \makeatletter
\def\maketitle{
\@author@finish
\title@column\titleblock@produce
\suppressfloats[t]}
\makeatother
 \usepackage{ragged2e}
\usepackage{amsmath}
\usepackage{amssymb}
\usepackage{physics}
\usepackage[normalem]{ulem}
\usepackage{graphicx}%
 \usepackage{setspace}
 \usepackage[skip=5pt]{caption}
\usepackage{dcolumn}
\usepackage{bm}
\usepackage[dvipsnames]{xcolor}

\begin{document}


\title{Correlation Effects on Coupled Electronic and Structural Properties of Doped Rare-Earth Trihydrides}


\author{Adam Denchfield}
\affiliation{Department of Physics, University of Illinois Chicago, Chicago, Illinois 60607, USA}
\author{Hyeondeok Shin}%
\affiliation{Computational Science Division, Argonne National Laboratory, Argonne, Illinois 60439, USA}

\author{Panchapakesan Ganesh}
\thanks{This author's contribution to the manuscript has been sponsored by UT-Battelle, LLC under Contract No. DE-AC05-00OR22725 with the U.S. Department of Energy. The United States Government retains and the publisher, by accepting the article for publication, acknowledges that the United States Government retains a non-exclusive, paid-up, irrevocable, worldwide license to publish or reproduce the published form of this manuscript, or allow others to do so, for United States Government purposes. The Department of Energy will provide public access to these results of federally sponsored research in accordance with the DOE Public Access Plan (\url{https://www.energy.gov/doe-public-access-plan}).}
\affiliation{Center for Nanophase Materials Sciences, Oak Ridge National Laboratory, Oak Ridge, Tennessee 37830, USA}
\author{Russell J. Hemley}
\affiliation{Department of Physics, University of Illinois Chicago, Chicago, Illinois 60607, USA}
\affiliation{Department of Chemistry, University of Illinois Chicago, Chicago, Illinois 60607, USA}
\affiliation{Department of Earth and Environmental Sciences, University of Illinois Chicago, Chicago, Illinois 60607, USA} 
\author{Hyowon Park}
\affiliation{Department of Physics, University of Illinois Chicago, Chicago, Illinois 60607, USA}
\affiliation{Materials Science Division, Argonne National Laboratory, Lemont, Illinois 60439, USA}



\date{\today}

\begin{abstract}
Rare-earth trihydrides ($R$H$_3$) exhibit intriguing coupled electronic and structural properties as a function of doping, hydrogen vacancies, and thermodynamic conditions. Theoretical studies of these materials typically rely on density functional theory (DFT), including the use of small supercells that may underestimate strong correlation effects and structural distortions {which in turn may influence their metallicity}. Here, we elucidate the roles of lattice distortions and correlation effects on the electronic properties of pristine and doped $R$H$_3$ by adopting DFT+U and Quantum Monte Carlo (QMC) methods. Linear-response constrained DFT (LR-cDFT) methods find Hubbard U $\approx 2$ eV for $R_d$ orbitals and U$\approx 6$ eV for {H$_s$/N$_p$} orbitals. The small U on Lu$_d$ orbitals is consistent with QMC calculations on LuH$_3$ and LuH$_{2.875}$N$_{0.125}$. In pure face-centered-cubic (FCC) $R$H$_3$ ($R$=Lu,Y), neither DFT nor DFT+U with the self-consistently determined U is enough to create a band gap, however a supercell with hydrogen distortions creates a small gap whose magnitude increases when performing DFT+U with self-consistently determined U values. 
Correlation effects, {in turn}, have a moderate influence on the coupled structural and electronic properties of doped RH$_3$ compounds {and may be important} when considering the competition between structural distortions and superconductivity. 
\end{abstract}


\maketitle


\section{Introduction}
\label{sec:intro}

Rare earth trihydrides remain enigmatic in a number of ways despite having been studied for over half a century. They are insulating with band gaps of 2-3 eV \cite{huiberts1996yttrium, kataoka2021face, van2001structural} but DFT typically finds them to be metallic \cite{alford2003first}. The addition of Hubbard U is unable to open a gap, even when solved with exact diagonalization \cite{wang1997nature}. The opening of a gap can be modeled with supercell distortions \cite{kelly1997theoretical, kerscher2012first}, by adding correlated hopping terms to a Hubbard model \cite{eder1997kondo}, or via a combination of supercell distortions and GW calculations \cite{nagara2012structural}. Under pressure,  $R$H$_{3}$ experiences multiple phases \cite{yao2010consecutive}, the gap closes \cite{ohmura2006infrared}, and shows significant vibrational anharmonicity at various pressures \cite{purans2021local, fukui2019characteristic}. If one ignores these correlation effects and structural distortions, DFT-based Eliashberg calculations indicate that FCC $R$H$_3$ should become superconducting at low pressures with $T_c \approx 20-50$ K \cite{durajski2014properties, kim2009predicted, lucrezi2024temperature}. Such T$_c$s are not possible since $R$H$_3$ are indeed insulating because of correlation effects and structural distortions, which must be taken into account for these materials in first-principles calculations. 

The oldest enduring mystery about $R$H$_3$ features the substoichiometric $R$H$_{3-\delta}$, which experience a metal-semiconductor-metal transition upon cooling \cite{libowitz1972temperature, shinar1990anomalous} at $\delta \approx 0.1-0.2$, thought to be concurrent with a quantum critical point \cite{hoekstra2003scaling}. Multiple transitions are observed in La[H,D]$_{3-\delta}$ at 230-270 K \cite{ito1983phase, kai1989heat} with at least one corresponding to a supercell distortion \cite{alikhanov1984neutron, udovic1999structural}. 
Clearly, RH$_{3-\delta}$ has a fascinating phase diagram marked by strong correlations, structural distortions, and significant anharmonicity. 

Given that the intriguing phase diagram of the cuprates originates in doping strongly correlated insulators, there is significant interest in studying doped/alloyed $R$H$_3$. For example, O-doped LaH$_{3-2x}$O$_x$ experiences an enhancement of H$^-$ conduction driven by significant anharmonicity and multi-hydrogen tunneling events \cite{fukui2019characteristic, fukui2022room}. Additionally, recent calculations found that hole-doping $R$H$_{3}$ would increase the stability of the cubic unit cell and lead to stronger electron-phonon coupling and superconducting critical temperatures up to 85 K \cite{villa2022superconductivity}. On that point, there is some evidence that N-doped LuH$_3$ exhibits superconductivity above 250 K below 20 kilobar \cite{salke2023evidence} and a number of first-principles structure searches to identify a possible stoichiometry and structure explaining the findings \cite{hao2023first, hilleke2023structure, ferreira2023search, lu2023electron, sun2023effect, lucrezi2024temperature}. The largest Eliashberg T$_c$ of 50 K is obtained from the FCC LuH$_3$ structure which is stabilized at low pressure by quantum and thermal effects \cite{shao2021superconducting, wu2023investigations, lucrezi2024temperature}, which is 85 meV/atom above the convex hull \cite{sun2023effect}. Despite this prediction of superconductivity at lower pressures, even though cubic LuH$_{3-\delta}$ and LuH$_{3-\delta}$N$_\epsilon$ has been recovered to ambient conditions \cite{li2023transformation, li2024stabilization}, superconductivity has only been reported in LuH$_3$ over 100 GPa \cite{shao2021superconducting}. This discrepancy could be attributed to two factors: the possible hydrogen superlattice distortions which are not {experimentally} characterized, and {structural distortions enhanced by correlation effects beyond DFT-PBE}.

The {high-throughput structure} searches for high-temperature superconductors {typically miss these two factors as they} use density functional theory (DFT) methods \cite{kohn1965self}, typically limited to small unit cells and computationally inexpensive functionals such as LDA \cite{kohn1965self} or PBE \cite{perdew1996generalized}. These {static} functionals, {including hybrid functionals} \cite{henderson2011accurate} miss the so-called `derivative discontinuity' present in the exact DFT functional \cite{kronik2020piecewise} which can lead to significant errors regarding Mott physics and other electron transfer effects, even in simple hydrogen-based systems \cite{mori2014derivative}. Both such effects are often lumped together as `correlation effects'.

To accurately account for these effects one often requires post-DFT methods such as Dynamical Mean Field Theory (DMFT), Quantum Monte Carlo (QMC) or full configuration interaction (FCI) methods. The Hubbard-corrected DFT approach, also called DFT+U(+V), attempts to circumvent the usage of {these computationally intensive} methods with augmented Hubbard parameters, treated in a mean-field manner \cite{tolba2018dft+}. Though initially used in an ad-hoc fashion to match electronic or crystal structure to experiment, there are now methods to determine these additional interaction parameters from first-principles, such as the constrained DFT \cite{himmetoglu2014hubbard} and RPA \cite{chang2024downfolding} methods. Such calculations of the Hubbard parameters depend sensitively on mesh size, projector choice, pseudopotentials, DFT functional, and other computational parameters \cite{timrov2018hubbard, tolba2018dft+, timrov2021self}, leading to a lack of transferability. 

To reduce {the sensitivity on input parameters}, it is possible to calculate the Hubbard parameters appropriate for a given structure with a linear response constrained DFT (LR-cDFT) approach \cite{himmetoglu2014hubbard, timrov2018hubbard, floris2020hubbard}. The self-consistent approach involves one converging the calculation of Hubbard parameters for a given unit cell \cite{timrov2021self}, relaxing the unit cell using those parameters \cite{timrov2020pulay}, and repeating the calculation of Hubbard parameters until convergence. When done self-consistently, LR-cDFT+U(+V) improves upon DFT-PBE results significantly with regards to experimental band gaps \cite{kirchner2021extensive} but often overestimates lattice constants regardless of the DFT functional used \cite{himmetoglu2014hubbard, timrov2021self}. We note such Hubbard-corrected DFT approaches may be missing additional correlation effects, like correlated hopping terms in RH$_3$ to reflect increased (correlated) hopping with electron occupation of H$_s$ orbitals \cite{eder1997kondo, ng1999theory}. One way to check the accuracy of the computed Hubbard parameters and resulting DFT+U calculations is through benchmarking against QMC calculations, which we do using fixed-node diffusion Monte Carlo (DMC) \cite{foulkes01}. 

Given the above considerations, it is imperative to address both lattice {distortions} and correlation effects in the screening of superconductors during a structure search, {including how the equilibrium crystal structure is changed due to correlation effects}. The above RH$_3$ examples illustrate discrepancies where DFT methods on small high-symmetry unit cells predict superconductivity where there is none {experimentally}. {However, an unexplored avenue is whether there are materials with large superconducting T$_c$ due to correlation effects and structural distortions that are filtered out in {high-throughput} structure searches}. Our initial work on large quaternary hydride structures indicates that some materials experience energy-lowering supercell distortions which could preserve or enhance T$_c$ \cite{denchfield2024designing}, and we note DMFT studies of superhydrides indicate that correlation effects on the Eliashberg spectral function do lead to increased T$_c$ \cite{wei2022exploring, wei2022high}. Therefore, we find it plausible that while {while correlation effects and structural distortions destroy superconductivity in $R$H$_{3}$, these factors may in fact enhance superconductivity when doping $R$H$_3$}.

\begin{figure}[t!]
  \centering
  \includegraphics[width=1.0\linewidth]{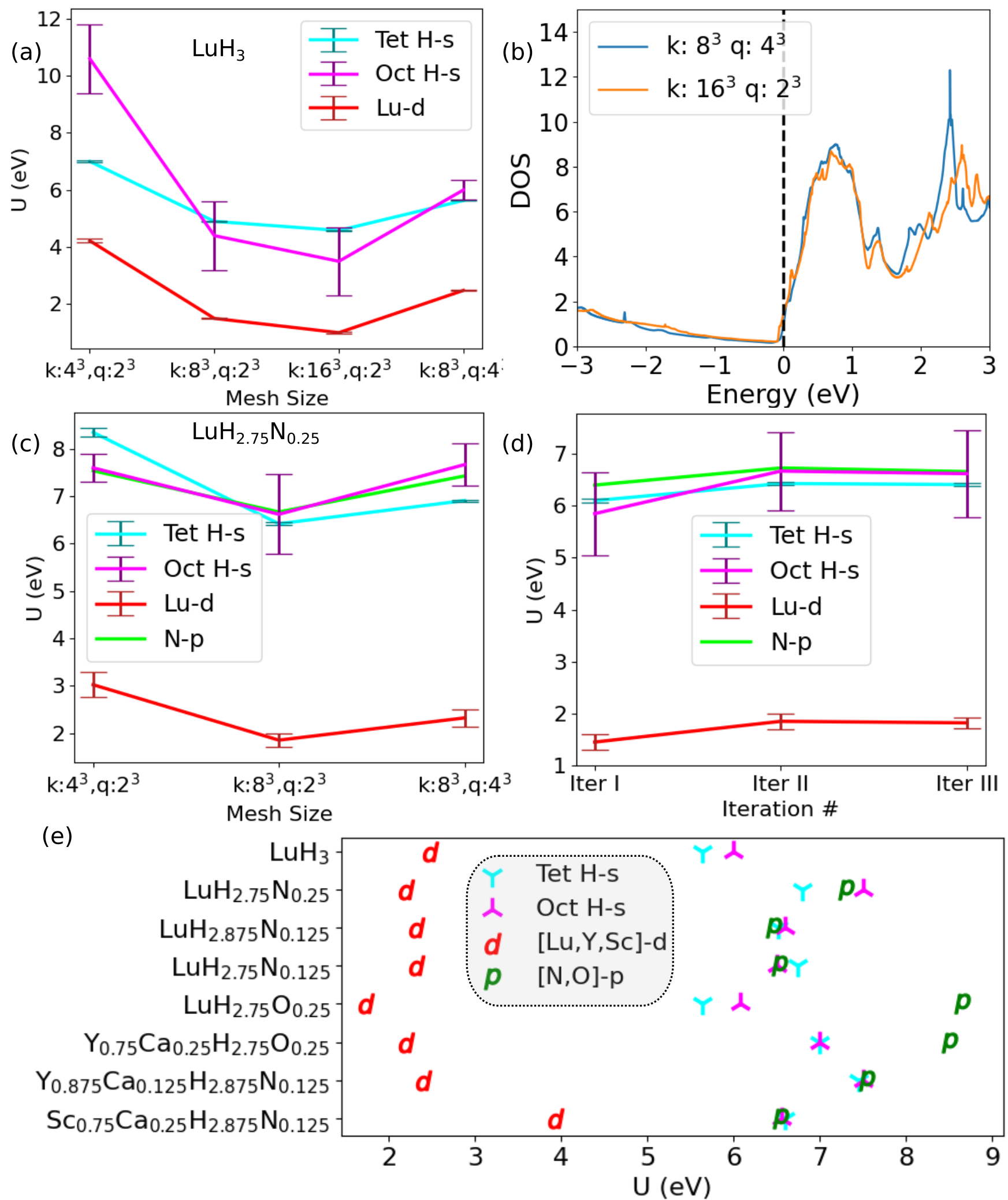}
  \caption{\justifying Convergence tests and evaluation of the Hubbard parameters for doped RH$_3$ hydrides using LR-cDFT. Error bars indicate the range of parameters computed for each specified orbital in the unit cell. (a) Dependence of the U parameters in FCC LuH$_3$ on k/q mesh size. (b) The DOS of FCC LuH$_3$ using the Hubbard parameters from two different mesh sizes. (c) Dependence of the U parameters in Pm3m LuH$_{2.75}$N$_{0.25}$ on k/q mesh size. (d) Convergence of Hubbard parameters in LuH$_{2.75}$N$_{0.25}$ upon self-consistent iteration of the LR-cDFT calculation. (e) Evaluation of Hubbard parameters for a variety of doped RH$_3$ compounds.}
  \label{fig:HP_U}
\end{figure}

\section{Methods}


\subsection{DFT+U}

In this work we use the DFT+U approach using self-consistently computed U parameters \sout{using} within the LR-cDFT approach \cite{himmetoglu2014hubbard, timrov2018hubbard, floris2020hubbard} as implemented in the \texttt{HP} code \cite{timrov2022hp} packaged in Quantum Espresso \cite{giannozzi2017advanced}. We used `ortho-atomic' projectors as the basis functions. We find for the systems studied that the computed V parameters are $\leq 1$ eV, so we only report DFT+U calculations and parameters. 

The k-point meshes determine the sampling of the Brillouin zone for determining electronic properties and the q-point meshes correspond to the periodicities of the external perturbations on certain atoms for the determination of Hubbard parameters. While the Hubbard parameters computed using the LR-cDFT have a nontrivial mesh-dependence [Fig.\ \ref{fig:HP_U}(a,c], there appears to be reasonable convergence. Furthermore, the electronic properties calculated using those parameters are not appreciably different for FCC LuH$_3$ [Fig.\ \ref{fig:HP_U}(b)] or LuH$_{2.75}$N$_{0.25}$ (not shown). 

For structural properties, we use the \texttt{phonopy} software \cite{togo2023first} to generate displacements, compute force constants, and calculate dynamical matrices using the acoustic sum rule \cite{brout1959sum}. The phonons computed at the $\Gamma$-point with our large 128-atom supercells should be reasonably representative of the structure's vibrational dynamics, including structural instabilities. The SSSP database \cite{prandini2018precision} was used to vet pseudopotentials for convergence of the computed pressure, phonon frequencies, and formation energies. Accordingly, we chose the pseudopotential for Lu from Ref. \cite{topsakal2014accurate}, and the \texttt{pslibrary} pseudopotentials for N and H \cite{dal2014pseudopotentials}. Each relaxed structure had remaining forces under \texttt{1e-5} Ry/Bohr, and energy differences were under \texttt{1e-5} Ry. We used recommended plane-wave cutoffs for the wavefunction (charge density) for each pseudopotential, which ranged from 75 (300) Ry for Lu-based hydrides to 90 (360) Ry for Sc-based hydrides. We generally used k-meshes corresponding to 0.13 \AA$^{-1}$ but performed convergence tests as necessary. We used \texttt{Quantum Espresso}'s default atomic projections for the projected density of states (PDOS) calculations. We used the improved tetrahedron method \cite{kawamura2014improved} for PDOS, electron localization function (ELF) \cite{savin1997elf}, and integrated local density of states (ILDOS) calculations. For structural relaxations and phonopy calculations we used smeared k-sums with a smearing value of 7e-3 Ry for speed and memory reasons, though we verified that usage of the tetrahedron method did not lead to different results for two examples. 

\subsection{QMC}

In this study, we employed fixed-node diffusion Monte Carlo method (DMC) \cite{foulkes01} as implemented with QMCPACK package \cite{QMCPACK} in order to estimate exact ground state properties of LuH$_3$ and N-doped LuH$_3$. Single Slater-determinant wavefunction using plane-wave basis set with Jastrow correlation coefficients were used as trial wavefunctions in the QMC algorithm, with up to three-body Jastrow factors were included in order to incorporate electron-ion, electron-electron, and electron-electron-ion correlations. Kohn-Sham orbitals in the Slater determinant were generated using PBE parametrization of the GGA exchange-correlation functional. In order to take on-site Coulomb interaction into account on $d$ orbitals on Lu and $p$ orbitals on N atom, we added Hubbard correction U in the trial wavefunction. Norm-conserving pseudopotentials from core correlation-consistent effective-core potentials (ccECPs) \cite{bennett17,annaberdiyev18} were used in this QMC study with 400 Ry kinetic energy cut-off. DMC calculations were done using 0.005~Ha$^{-1}$ time steps within the non-local $T$-move approximation. \cite{casula10} In order to reduce one-body finite-size effects from the periodic boundary conditions applied in the DMC calculations, we employed twist-averaged boundary conditions \cite{lin01} with up to total 64 twists on LuH$_3$ and LuH$_{2.875}$N$_{0.125}$ supercell. The model periodic Coulomb (MPC) interaction \cite{drummond08} and Chiesa-Ceperley-Martin-Holzmann kinetic energy correction \cite{chiesa06} were additionally applied to minimize the finite size effect.

\section{Results and Discussion}


\subsection{LR-cDFT calculations of Hubbard U}

The computed Hubbard parameters for doped RH$_3$ {do not vary much across different compounds except} an increase in U$_d$ in Sc-based compounds [Fig.\ \ref{fig:HP_U}(e)]. The magnitude of the computed Hubbard parameters for R$_d$, N$_p$, and O$_p$ are in reasonable agreement with the Hubbard parameters computed in a variety of other systems using a similar methodology \cite{kirchner2021extensive}. Such large Hubbard parameters for H$_s$ orbitals is unexpected, but strong correlation effects for the H$_s$ orbitals in YH$_3$ and LaH$_3$ do successfully explain their electronic properties \cite{eder1997kondo, ng1999theory}. This is due to weakly bonded hydrogen presenting a case where the derivative discontinuity error in DFT functionals is especially egregious \cite{mori2014derivative}.  DFT+U provides approximate solutions to these issues at a mean-field level, which justifies the high U$_{H_s}$ values. 

\subsection{LuH$_3$}

To ascertain whether FCC LuH$_3$ develops a band gap with correlation effects, we have performed the LR-cDFT method as well as a Hubbard U scan using DMC (Fig.\ \ref{fig:QMC_LuH3}) to compute Hubbard parameters for FCC LuH$_3$. We found the LR-cDFT results depend somewhat on the size of k-point/q-point mesh used [Fig.\ \ref{fig:HP_U}(a)] but the resulting DFT+U density of states (DOS) does not significantly change even when the computed U parameters change by over an eV [Fig.\ \ref{fig:HP_U}(b)]. The self-consistent DFT+U does not open a band gap, even with volume relaxation with U; increasing U$_{Lu_d}$ to 7.5 eV also did not result in a band gap opening. Usage of hybrid functionals also does not result in a gap for FCC LuH$_3$ \cite{wu2023investigations}.

\begin{figure}[h!]
  \centering
  \includegraphics[width=1.0\linewidth]{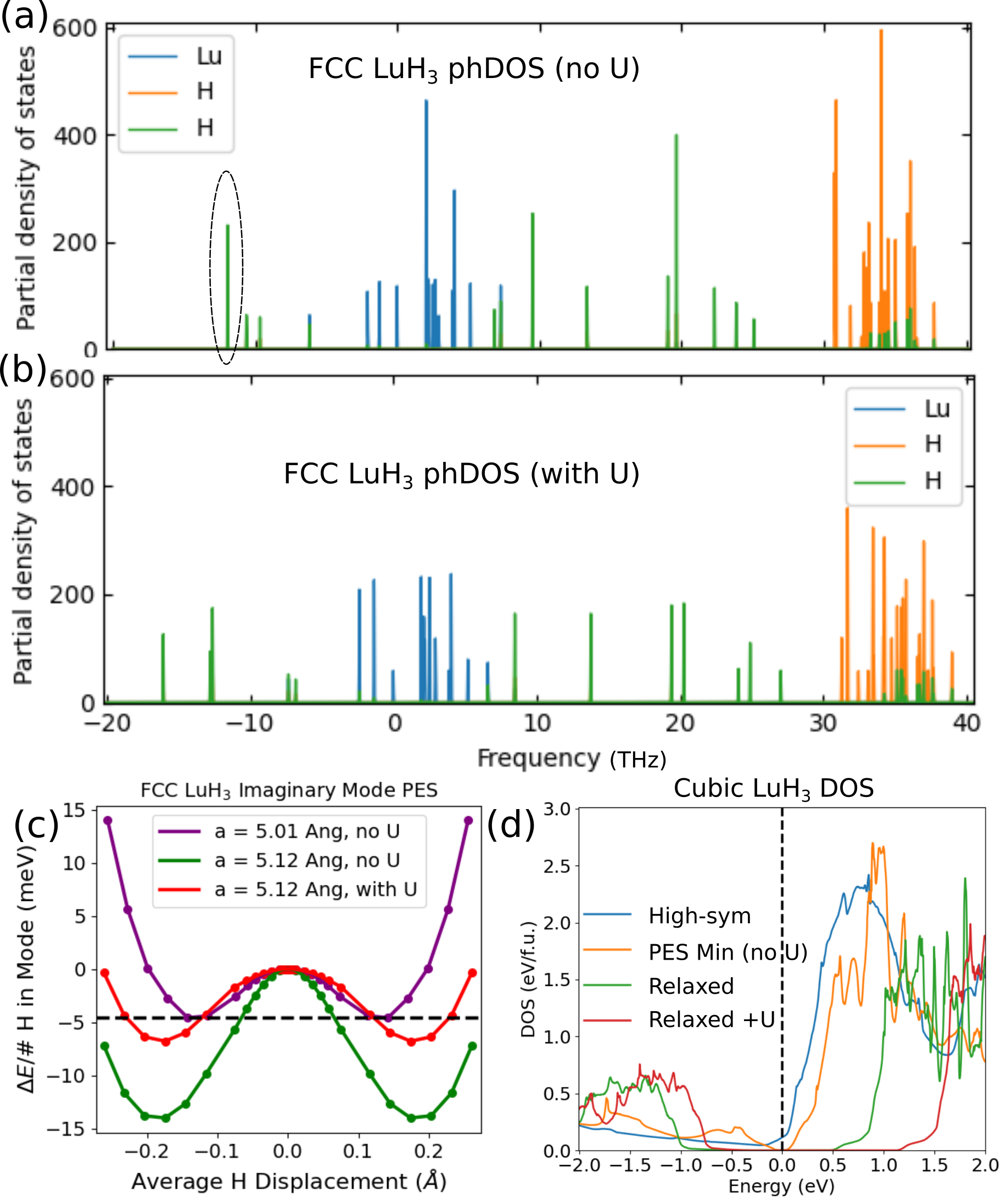}
  \caption{\justifying Electronic and structural properties of FCC LuH$_3$. (a) The projected phonon density of states (phDOS) of a 128-atom supercell of FCC LuH$_3$ at $\Gamma$ computed without U with $a =$ 10 \AA. Imaginary frequencies are plotted as negative. (b) The phDOS of the same supercell as (a) computed with DFT+U. (c) PES from DFT calculations on the LuH$_3$ supercell displaced along the eigenmode with the largest imaginary frequency. (d) The total DOS of LuH$_3$ for three structural configurations (high symmetry, PES minimum, fully relaxed from the PES minimum). The fully relaxed configuration also had the DOS computed with U-values computed from the constrained DFT method. }
  \label{fig:LuH3}
\end{figure}

For reference, pristine HCP LuH$_3$ is insulating with a non-reflective red color \cite{daou1988percolating, drulis2009low, moulding2023pressure} that persists in its cubic state \cite{moulding2023pressure, li2023transformation}, indicating its band gap is in the 1.8-2.4 eV range similar to YH$_3$ \cite{kataoka2021face}. By finding low-energy supercell distortions through a phonon analysis and performing a relaxation and band structure analysis with DFT+U, we are able to reproduce this band gap (Fig.\ \ref{fig:LuH3}). 

This is consistent with {the fact that} supercell distortions are needed to fully explain the band gap in YH$_3$ \cite{nagara2012structural}, we create a 128-atom cubic supercell of FCC LuH$_3$ with a=10.04 \AA\ and relax it from its high symmetry FCC phase, which due to a symmetry-induced cancellation of forces does not lead to a broken-symmetry structure [Fig. \ref{fig:LuH3_distortions}(a)]. To {study its structural stability and possibly} find a lower-energy structure, we compute its classical $\Gamma$-point phonon spectrum with \texttt{phonopy} \cite{togo2023first}, whose harmonic frequencies are shown in Fig.\ \ref{fig:LuH3}(c). Some of these frequencies are imaginary, indicating a classical structural instability. The usage of DFT+U using Hubbard parameters derived from LR-cDFT yields a similar set of phonons, with small changes to the computed frequencies [Fig.\ \ref{fig:LuH3}(d)]. We compute the DFT-based potential energy surface (PES) of the system displaced along the mode with the largest imaginary frequency [Fig.\ \ref{fig:LuH3}(c)], with the displacement visualized in Fig.\ \ref{fig:LuH3_distortions}(b).

While there are indications that quantum nuclear and thermal effects can stabilize the high-symmetry FCC state at low pressures and high enough temperatures \cite{lucrezi2024temperature}, we find correlation effects to expand the lattice and alter the PES. At a fixed lattice constant of 5.12 \AA\ (the lower bound on FCC LuH$_3$'s experimental lattice constant \cite{tkacz2007pressure}), the displacement lowers the energy of the system by 14 meV/hydrogen participating in the mode {with PBE}, but the energy is lowered much less with {PBE}+U [Fig.\ \ref{fig:LuH3}(c)]. The total DOS indicates that the hydrogen sublattice distortion opens a gap, and subsequent relaxation at both a PBE and PBE+U level further distorts the lattice and leads to a optical gap of 1.4 and 1.9 eV, respectively [Fig.\ \ref{fig:LuH3_distortions}(c)]. The latter gap is consistent with the red color observed for LuH$_3$ \cite{moulding2023pressure, li2023transformation}. We visualize the three LuH$_3$ structures in Fig.\ \ref{fig:LuH3_distortions}. The relaxed structure is a local minimum of the PES since the global energy minimum is a distorted HCP structure \cite{schollhammer2017first}. This minimum is 56 meV/atom lower in energy than the high symmetry FCC structure, and 50.9 meV/atom lower in energy than the structure corresponding to the PES minimum. For reference, trigonal LuH$_3$ is 85 meV/atom lower in energy than FCC LuH$_3$ \cite{sun2023effect}. Regardless, cubic LuH$_{3-\delta}N_\epsilon$ can be recovered to ambient conditions \cite{li2024stabilization} and pressurization of trigonal LuH$_3$ to 1.9 GPa led to a phase consistent with a distorted cubic supercell of LuH$_3$ \cite{moulding2023pressure}. Based on our computed electronic and structural properties of cubic LuH$_3$, the presence of correlation effects encourages distortions into an insulating state {that has} a gap consistent with that seen experimentally.





\subsection{LuH$_{2.875}$N$_{0.125}$}
\begin{figure}
  \centering
  \includegraphics[width=1.0\linewidth]{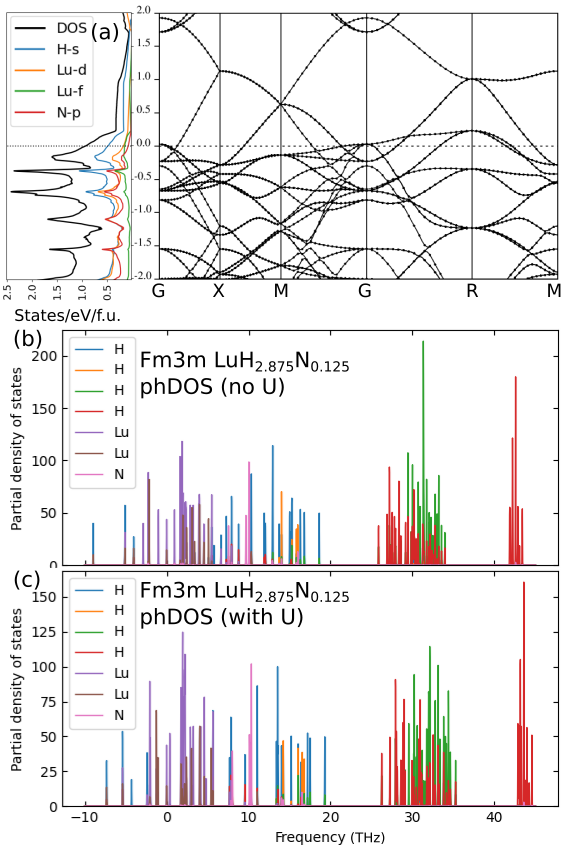}
  \caption{\justifying Electronic and structural properties of Fm$\overline{3}$m LuH$_{2.875}$N$_{0.125}$. (a) Electronic PDOS and band structure with DFT+U, (b) Phonon frequencies with a=5.07 \AA\ using PBE, (c) Phonon frequencies with a=5.07 \AA\ using PBE+U. Hubbard parameters used were derived using the LR-cDFT method (see Fig.\ \ref{fig:HP_U}).}
  \label{fig:Lu8H23N}
\end{figure}

While our previous work on LuH$_{2.875}$N$_{0.125}$ \cite{denchfield2024electronic} suggests correlation effects on Lu$_d$ orbitals are capable of bringing narrow hydrogen bands close to E$_F$ and further flattening them, the value of U used was not {obtained} from first principles, which this work will now address. Furthermore, while we found nuclear quantum effects (NQE) aided the dynamical stability of Fm$\overline{3}$m LuH$_{2.875}$N$_{0.125}$ \cite{denchfield2024quantum}, we hinted that correlation effects could alter this picture indirectly through lattice expansion. This is supported by the observation that FCC LuH$_{3-\delta}$N$_{0.166}$ (1.3\% N by weight) has been synthesized and recovered to ambient conditions with a lattice constant of 5.156 \AA\ \cite{li2024stabilization}, corresponding to a 3\% larger lattice than studied for Fm$\overline{3}$m LuH$_{2.875}$N$_{0.125}$ in \cite{denchfield2024quantum}. 

We find that the Hubbard parameters {for Lu$_d$ orbitals} computed via the LR-cDFT method {are smaller than} the U$_{Lu_d} = 5.5, 8.2$ eV used in our previous studies \cite{denchfield2024electronic, denchfield2024quantum} [Fig.\ \ref{fig:HP_U}(e)]. As self-consistent PBE+U with LR-cDFT leads to overestimated lattice constants \cite{timrov2018hubbard, timrov2021self}, we relax the system with PBE+U under a simulated pressure of 5 GPa, leading to a lattice constant of 10.14 \AA, which still exceeds the PBE value of 10.04 \AA\ under no simulated pressure. We find some hydrogen states below E$_F$ have their energies lowered relative to the PBE values (see Ref.\ \cite{denchfield2024electronic}) by the Hubbard parameters [Fig.\ \ref{fig:Lu8H23N}(a)]. To estimate the change in stability properties at a fixed lattice constant, we compare the $\Gamma$-point phonon frequencies and plot them as a discrete phonon density of states (phDOS) with PBE [Fig.\ \ref{fig:Lu8H23N}(b)] and PBE+U [Fig.\ \ref{fig:Lu8H23N}(c)]. {Most of the} phonon frequencies change by 1-2 THz, slightly stiffening the lattice with PBE+U. While there are phonons with imaginary frequencies (representing a lattice instability), it is expected nuclear quantum  effects (NQE) will significantly stabilize the lattice as it did at a=10.04\ \AA\ at the PBE level \cite{denchfield2024quantum}. 

\begin{figure}
  \centering
  \includegraphics[width=1.0\linewidth]{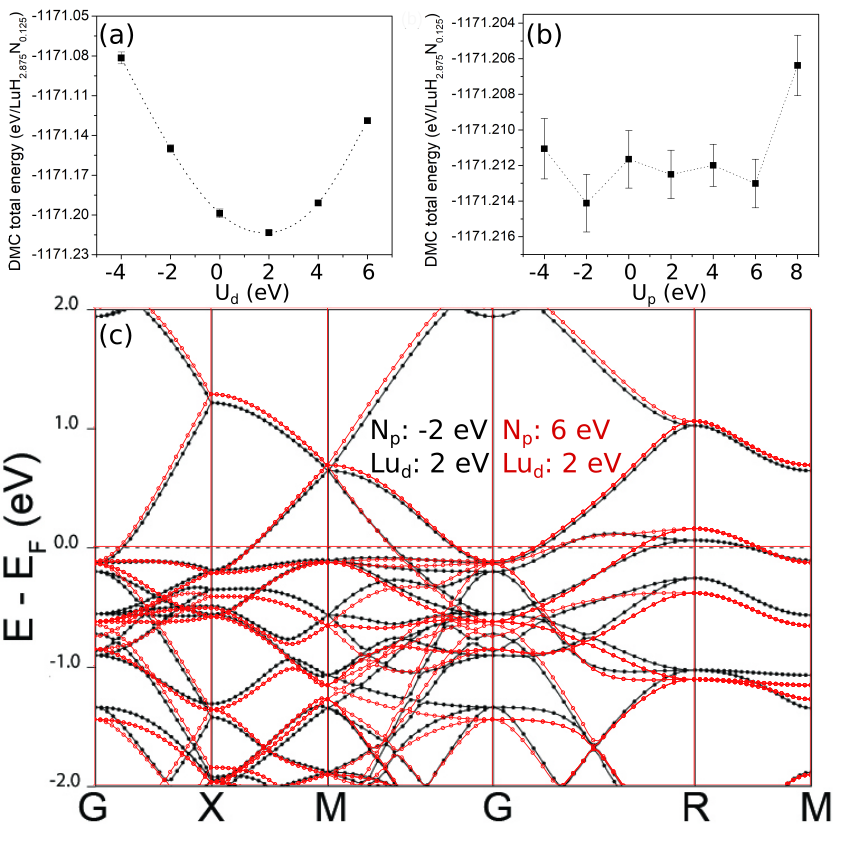}
  \caption{DMC total energy as function of U in PBE+U trial wavefunction for (a) 5$d$ orbital of Lu atom and (b) 2$p$ orbital of N atom on LuH$_{2.875}$N$_{0.125}$. (c) Comparison of DFT+U band structures using U$_{Lu_d} = 2$ eV and U$_{N_p} = \{-2,6\}$ eV, {with ccECPs}.}
  \label{fig:DMC_U}
\end{figure}

In order to estimate the optimal value of U for 5$d$ orbital on Lu atom and 2$p$ one on N atom within DMC framework by minimizing DMC total energy, we performed DMC with different values of U in PBE+U trial wavefunction for 464 electrons LuH$_{2.875}$N$_{0.125}$ supercell consisting of total 32 formula units. We are unfortunately unable to perform the same analysis for H-1s orbitals due to the nodeless 1s wavefunctions \cite{dubecky2013quantum}.
As seen in Figure~\ref{fig:DMC_U}(a), non-negligible energy difference of 0.02(1) eV/f.u. is shown between DMC energy estimated using U$_{d}$ = 0 and U$_{d}$ = 2.0 eV in PBE+U trial wavefunction.  
Using quadratic fit, the optimal value of U$_d$ for LuH$_{2.875}$N$_{0.125}$ was estimated to be 1.9(2) eV, respectively, which are close to the Hubbard parameters obtained using LR-cDFT method for LuH$_{3}$ or LuH$_{2.875}$N$_{0.125}$.
Figure~\ref{fig:DMC_U}(b) displays DMC total energy of LuH$_{2.875}$N$_{0.125}$ at different values of U$_{p}$ on 2$p$ orbitals in N atom with U$_{d}$ fixed as 2.0 eV.  
The U$_{p}$ dependency on DMC energy has a much different trend compared to U$_{d}$. 
Rather than parabolic, the DMC energy is almost uniform in the range of U$_{p}$ = 0.0 - 6.0 eV with statistical errors of 1 - 2 meV/LuH$_{2.875}$N$_{0.125}$ and local minima at -2 eV and 6 eV. While the statistical errors are small, the DMC energy differences are small enough to be comparable. Given that the energy difference between the two N-2$p$ minima are smaller than the DMC statistical error, we find the LR-cDFT U$_{N-2p}$ values to be consistent with the DMC results.  
Based on our DMC results for LuH$_{2.875}$N$_{0.125}$, we can assume that the effect of U$_{p}$ on the nodal surface of the PBE+U wavefunction is much smaller than that of U$_{d}$.

For further investigation of the effect of U$_{p}$ on electronic and optical properties of LuH$_{2.875}$N$_{0.125}$, we computed PBE+U projected density states and band structure using ccECPs.
In Figure~\ref{fig:DMC_U}(c), we see narrow valence bands near the Fermi level, indicating strong electron coupling with the addition of on-site Coulomb interaction on Lu-$5$d and N-2$p$ states.
The corresponding PDOS (Fig.\ \ref{fig:PDOS_ccECP}) confirms there are many N-2$p$ states at E$_F$, {but the variations due to U$_{N-2p}$ are not significant, explaining} the weak DMC energy dependence on U$_{N-2p}$ unexpected. 

While our DFT+U calculations do not indicate band gap opening in FCC LuH$_{2.875}$N$_{0.125}$, studies of YH$_3$ indicate the band gap can be opened with the consideration of non-density-density interactions \cite{eder1997kondo, nagara2012structural}.
In order to confirm that FCC LuH$_{2.875}$N$_{0.125}$ is metallic with an accurate electronic structure probe, we compute the DMC fundamental gap $E_{qp}$ using the equation $E_{qp} = E(N+1) + E(N-1) - 2E(N)$, where $E(N)$, $E(N+1)$, and $E(N-1)$ represent ground-state total energy of $N$, $N+1$, and $N-1$ electrons system, respectively. The DMC fundamental gap of FCC LuH$_{2.875}$N$_{0.125}$ was estimated using $N$ = 464 electrons {and confirmed} that it is gapless. 

While the DMC calculations rule out a Mott insulating ground state, superconducting or structural instabilities may occur. Despite the results in Fig. \ref{fig:Lu8H23N} implying a structural instability, the consideration of quantum nuclear effects predicts structural (meta)stability of FCC LuH$_{2.875}$N$_{0.125}$ at 2 GPa  \cite{denchfield2024quantum}. {At ambient pressure, it is possible the structure distorts into an insulating state (see SM in Ref.\ \cite{denchfield2024quantum}).}




\subsection{Y$_{0.875}$Ca$_{0.125}$H$_{2.875}$N$_{0.125}$ and Sc$_{0.75}$Ca$_{0.25}$H$_{2.875}$N$_{0.125}$}

\begin{figure}
  \centering
  \includegraphics[width=1.0\linewidth]{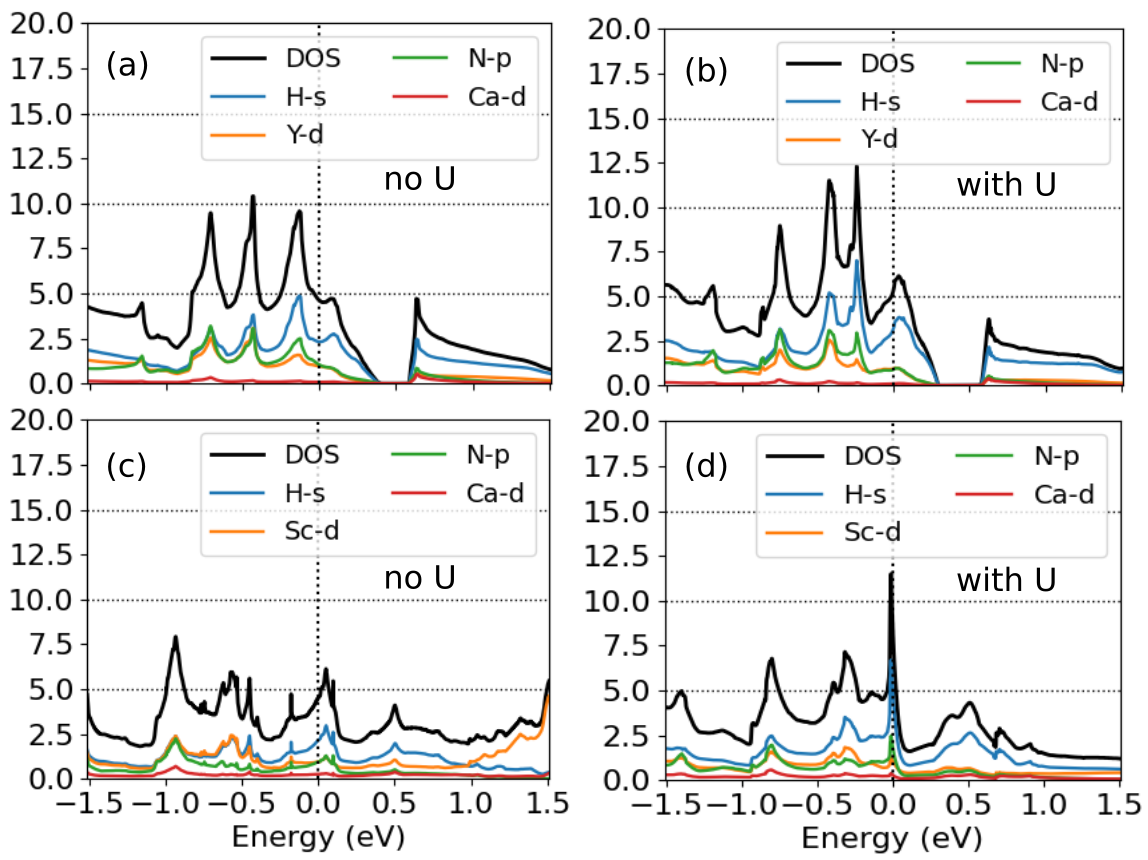}
  \caption{\justifying Comparison of the PDOS of quaternary hydrides with and without self-consistently computed Hubbard U parameters. (a) F$\overline{4}$3m Y$_{0.875}$Ca$_{0.125}$H$_{2.875}$N$_{0.125}$ at 1 GPa (a=5.12 \AA) without U, (b) with U (a=5.20 \AA). (c) Fm$\overline{3}$m Sc$_{0.75}$Ca$_{0.25}$H$_{2.875}$N$_{0.125}$ at 20 GPa without U (a=4.62 \AA), (d) with U (a=4.7 \AA). }
  \label{fig:ScY}
\end{figure}

To illustrate the effect of the correlation effects in other doped FCC RH$_3$ structures, we address quaternary hydrides based on FCC RH$_3$ \cite{denchfield2024designing}. We find F$\overline{4}$3m Y$_{0.875}$Ca$_{0.125}$H$_{2.875}$N$_{0.125}$ at 1 GPa has a predicted lattice expansion from a=10.24\AA\ to a=10.40\AA\ when including Hubbard parameters in the relaxation. Furthermore, the inclusion of Hubbard parameters concentrates spectral weight at E$_F$ [Fig.\ \ref{fig:ScY}(a-b)], likely enhancing a possible superconducting T$_c$. A similar lattice expansion from a=9.24\AA\ to a=9.40\AA\ is observed in Fm$\overline{3}$m Sc$_{0.75}$Ca$_{0.25}$H$_{2.875}$N$_{0.125}$ at 20 GPa. The accumulation of spectral weight at E$_F$ is more prominent in this example [Fig.\ \ref{fig:ScY}(c-d)].

While the previous LuH$_{2.875}$N$_{0.125}$ example addressed the effect of correlation effects on lattice stability at a fixed lattice constant, we also address the effect of correlation-induced volume expansion on the phonon frequencies in Fm$\overline{3}$m Sc$_{0.75}$Ca$_{0.25}$H$_{2.875}$N$_{0.125}$ in Figure \ref{fig:Sc_phonon}. Comparing the frequencies with and without U, it can be seen that the volume expansion leads to a less stable lattice, but the phonon frequencies at nearly equivalent lattice constants using PBE vs PBE+U show overall hardening in the PBE+U case. Based on calculations in our other work \cite{denchfield2024designing} we estimated the system is stabilized by NQE $\gtrapprox$ 30 GPa; the inclusion of correlation effects may increase the pressure needed to stabilize the Fm$\overline{3}$m structure by a few GPa. We remark that the distortion of the hydrogen atoms corresponding to the phonon with largest imaginary frequency was actually found to increase the estimated superconducting T$_c$ \cite{denchfield2024designing}, indicating a case where structural distortions may be desirable for superconductivity. 



\section{Conclusions}

We have estimated the role of correlation effects in pure and doped FCC RH$_3$ via calculation of Hubbard parameters from first principles. While correlation effects appear to directly affect electronic properties moderately, they induce volume expansion and often promote structural distortions, leading to an enhanced change in electronic properties. Hydrogen-based supercell distortions of FCC LuH$_3$ in conjunction with correlation effects at the level of self-consistent DFT+U give rise to a band gap consistent with that observed in YH$_3$ and LuH$_3$. 
Hubbard U scanning in FCC LuH$_{3}$ and Fm$\overline{3}$m LuH$_{2.875}$N$_{0.125}$ within a QMC scheme provides values of U on Lu-5$d$ and N-2$p$ orbitals consistent with the corresponding LR-cDFT results. Usage of DMC for computing the fundamental gap confirms that Fm$\overline{3}$m LuH$_{2.875}$N$_{0.125}$ is metallic (rather than a correlated insulator). Given the consistency of LC-cDFT and DMC methods for computing U and the fairly small Hubbard parameter changes for a variety of doped LuH$_3$ structures, it appears the Hubbard U is fairly transferable in these systems. 
We also confirm in two quaternary hydride examples that the introduction of self-consistently computed Hubbard parameters can appreciably change electronic properties at E$_F$ even without significant structural distortions. Overall, we find correlation effects in doped RH$_3$ to have the largest effects when considered in conjunction with structural deformations and are likely necessary considerations {when considering the competition between structural distortions and superconductivity} for these materials. 

\section{Acknowledgements}
Authors are grateful to Benjamin Kincaid, Haihan Zhou, Abdulgani Annaberdiyev, and Lubos Mitas for optimizing Lu ccECP. 
 This research was supported by the National Science Foundation (DMR-2104881, R.H.), the Department of Energy (DOE) National Nuclear Security Administration through the Chicago/DOE Alliance Center (DE-NA0003975; A.D., R.H.), the DOE Office of Science (DE-SC0020340, R.H.), and NSF SI2-SSE Grant 1740112 (H.P.). This research used resources of the National Energy Research Scientific Computing Center (NERSC), a U.S. Department of Energy Office of Science User Facility located at Lawrence Berkeley National Laboratory, operated under Contract No. DE-AC02-05CH11231 using NERSC award BES-ERCAP0023615.
H.S. and P.G. (QMC calculations) were supported by the U.S. Department of Energy, Office of Science, Basic Energy Sciences, Materials Sciences and Engineering Division, as part of the Computational Materials Sciences Program and Center for Predictive Simulation of Functional Materials. An award of computer time was provided by the Innovative and Novel Computational Impact on Theory and Experiment (INCITE) program. This research used resources of the Argonne Leadership Computing Facility, which is a DOE Office of Science User Facility supported under contract DE-AC02-06CH11357.

\section{Data Availability.} Inputs and output files for DFT and DMC are available at the Materials Data Facility \cite{blaiszik2016materials,blaiszik2019data} ({\tt https://doi.org/10.18126/3a3m-sm24}), DOI: 10.18126/3a3m-sm24.
 
\newpage

\bibliography{A.bib}


\appendix


\clearpage



\renewcommand{\thesection}{\Alph{section}}    

\renewcommand{\thefigure}{\Alph{section}\arabic{figure}}


\section{Distortions of FCC LuH$_3$}

\setcounter{figure}{0}

\begin{figure}[h]
  \centering
  \vspace{-0.2cm}
  \includegraphics[width=0.55\linewidth]{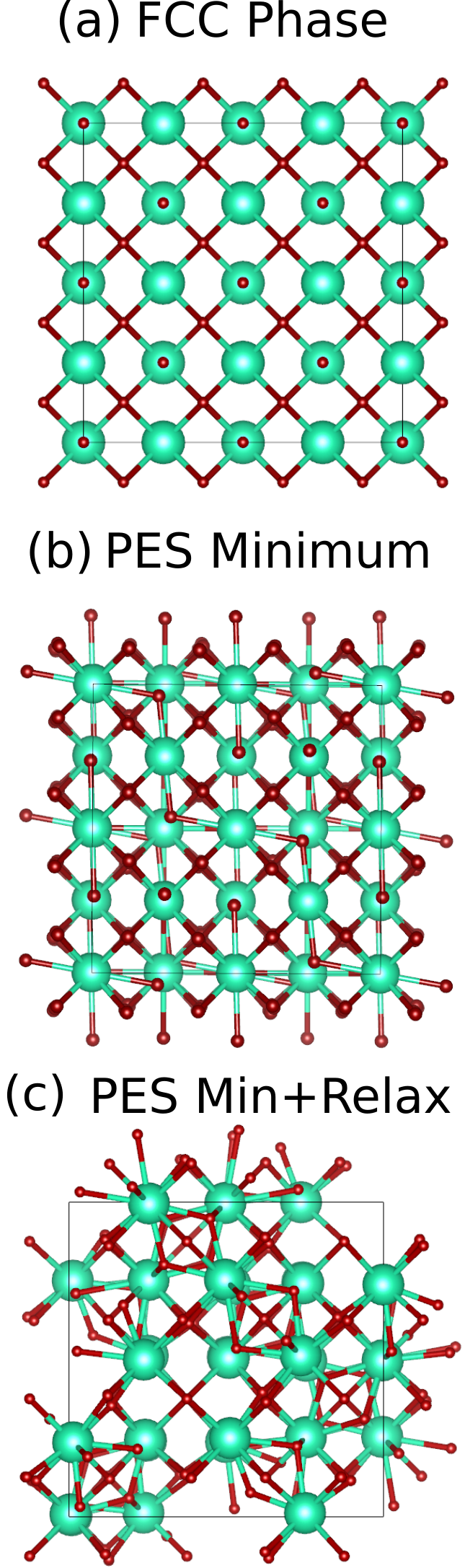}
  \caption{(a) The original high-symmetry supercell of FCC LuH$_3$. (b) The supercell with atomic displacements corresponding to the PES minimum. (c) A DFT relaxation of the supercell from the PES minimum, resulting in additional distortions of the cell.}
  \label{fig:LuH3_distortions}
\end{figure}

\section{QMC Analysis of FCC LuH$_3$}
\setcounter{figure}{0}
\begin{figure}[h]
  \centering
  \vspace{-0.7cm}
  \includegraphics[width=1.0\linewidth]{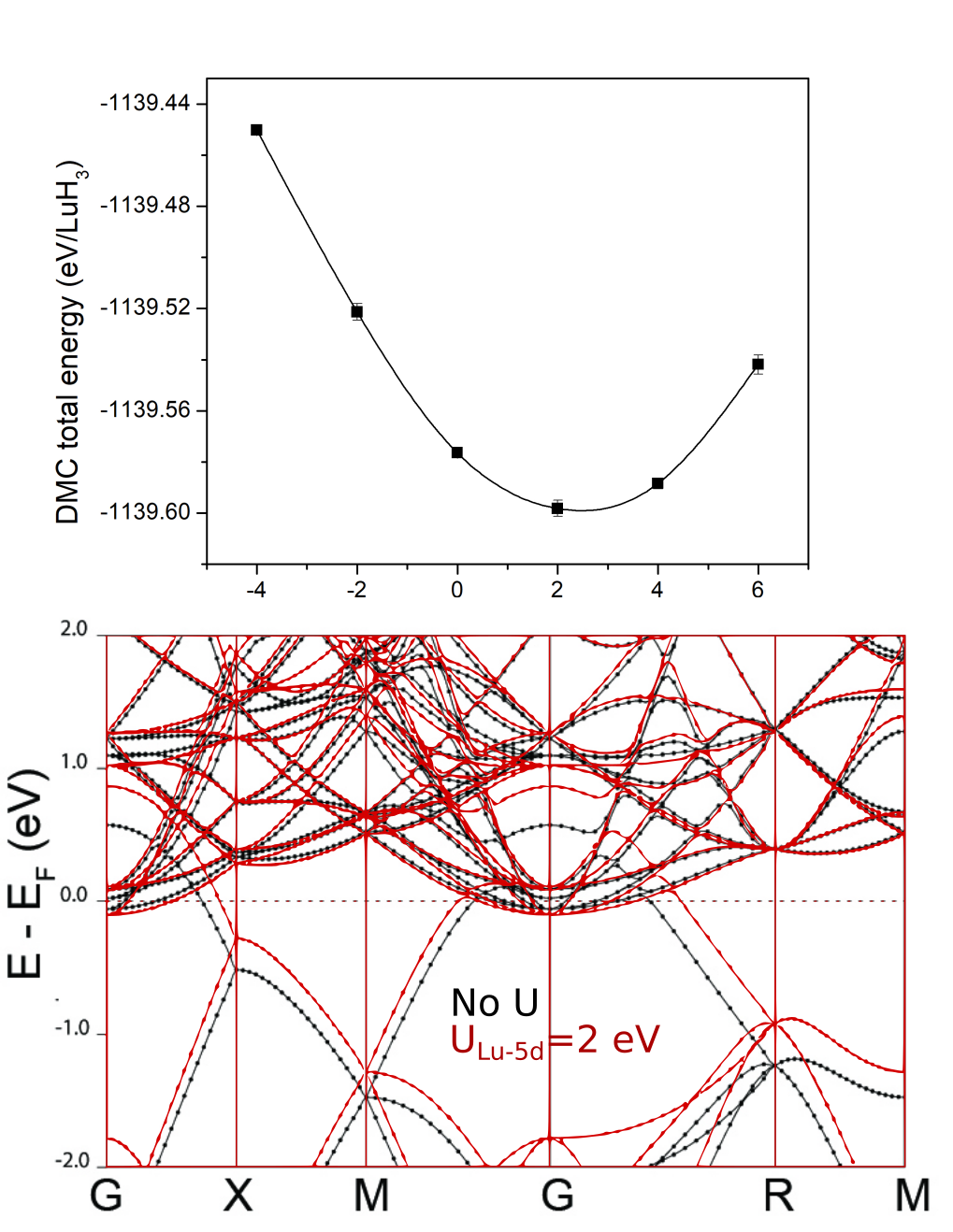}
  \caption{QMC Hubbard U scan (top) and PBE(+U) band structure of FCC LuH$_3$ with U$_{d}$ = 0 eV and U$_{d}$ = 2.0 eV (bottom).}
  \label{fig:QMC_LuH3}
\end{figure}

Figure \ref{fig:QMC_LuH3} illustrates agreement between the DMC scan for the U$_{Lu-5d}$ Hubbard parameter and the LR-cDFT result, with both yielding optimal values close to 2 eV. Also plotted are the band structures using PBE+U with the ccECP, illustrating that addition of the Hubbard U appears to encourage charge transfer from Lu-5$d$ to H-1$s$ orbitals. 

\section{DFT+U Dependence of Multicomponent Hydride Properties}
\setcounter{figure}{0}

\begin{figure}[h]
  \centering
  \includegraphics[width=1.0\linewidth]{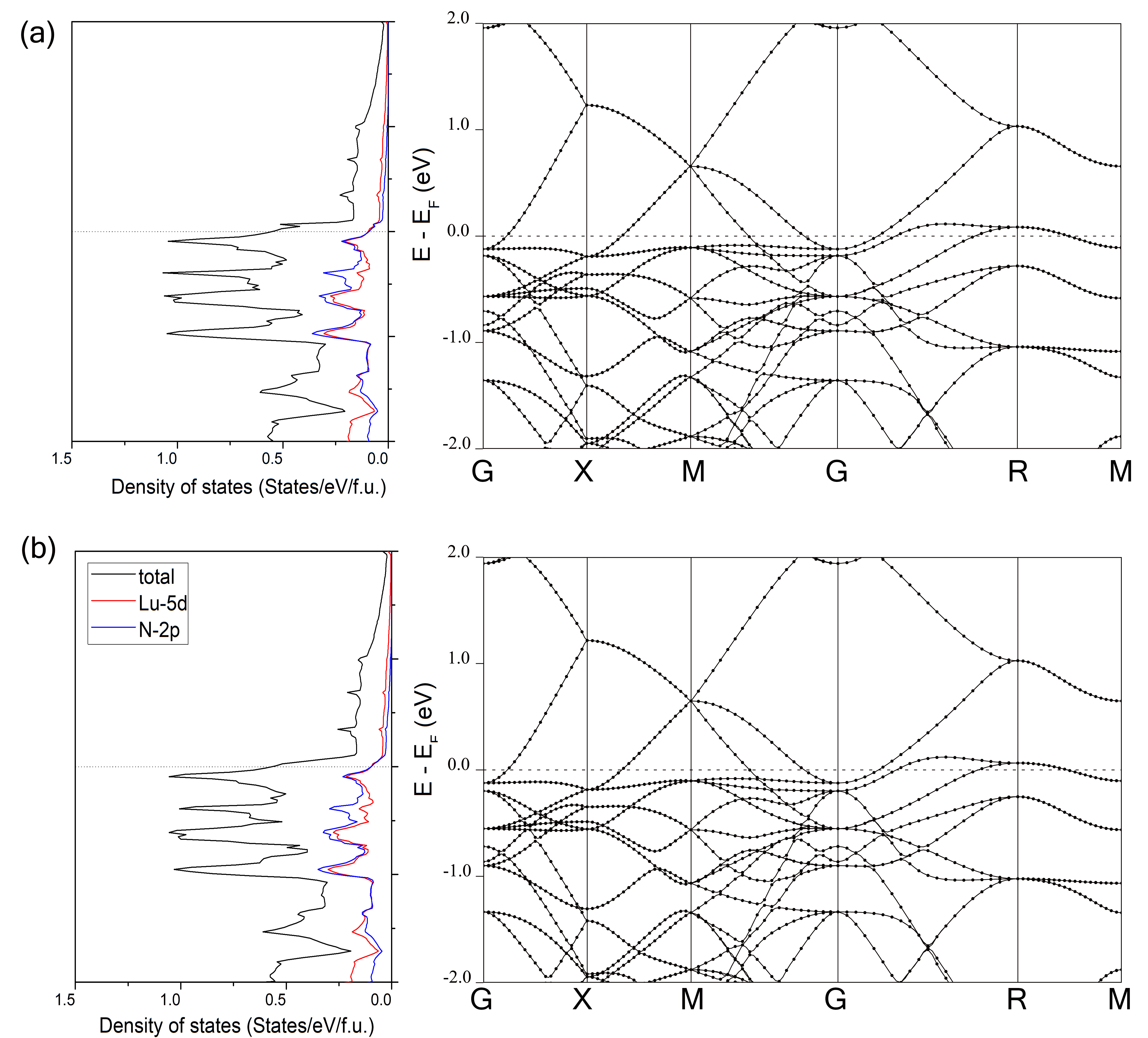}
  \caption{PBE+U projected density of states and band structure of LuH$_{2.875}$N$_{0.125}$ with (a) U$_{d}$ = 2.0 eV and (b) U$_{d}$ = 2.0 eV, U$_{p}$ = -2.0 eV.}
  \label{fig:PDOS_ccECP}
\end{figure}

Figure \ref{fig:PDOS_ccECP} illustrates the bandstructure and PDOS of Fm$\overline{3}$m LuH$_{2.875}$N$_{0.125}$ computed at the PBE+U level with U$_{Lu_d} = 2$ eV and different values of U$_{N_p}$ using the ccECPs used for the QMC calculations, illustrating the sharpness of the DOS near E$_F$.

\begin{figure}[h]
  \centering
  \includegraphics[width=1.0\linewidth]{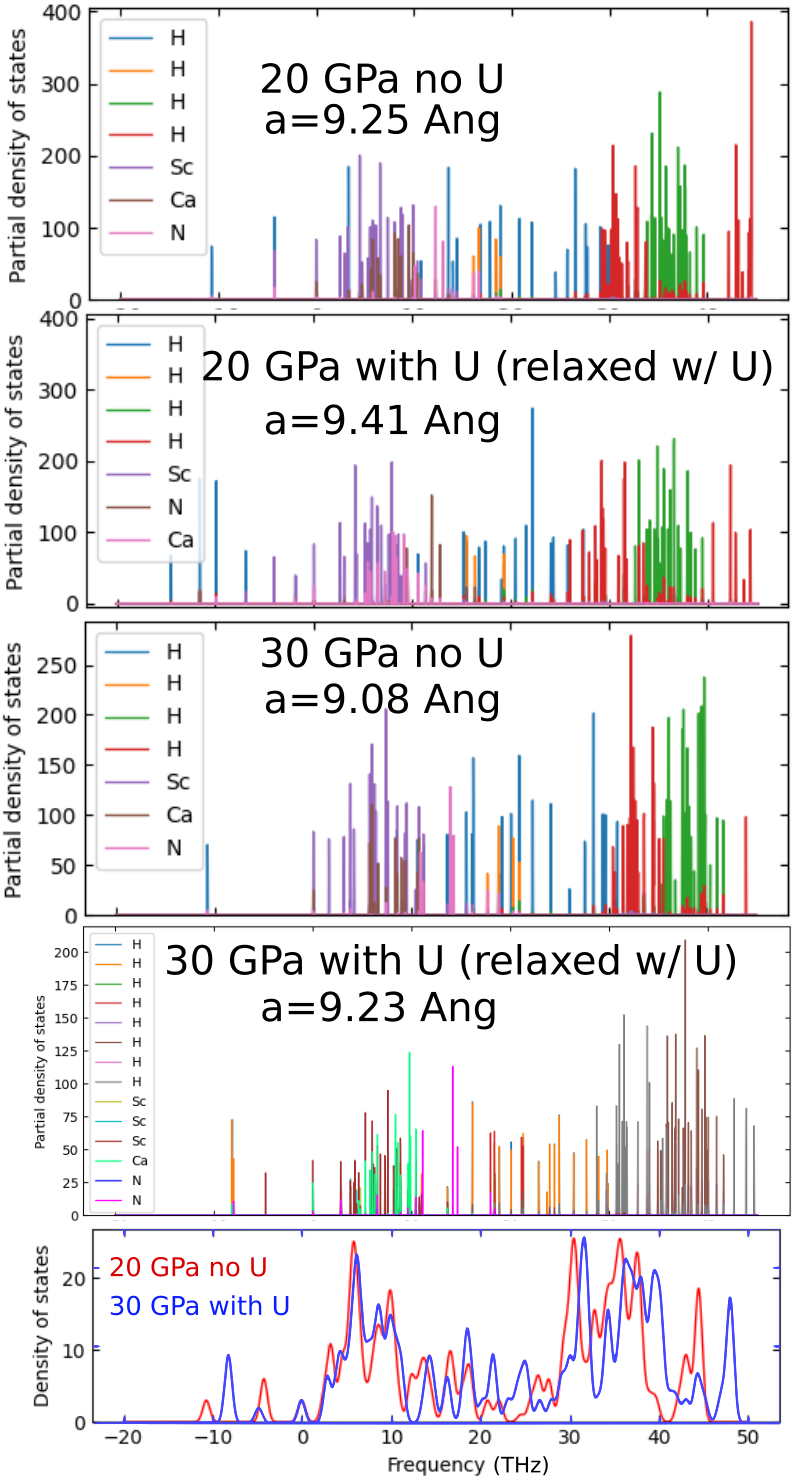}
  \caption{\justifying Comparison of phonon projected density of states (phDOS) of Fm$\overline{3}$m Sc$_{0.75}$Ca$_{0.25}$H$_{2.875}$N$_{0.125}$ at 20 GPa and 30 GPa without and with Hubbard parameters. }
  \label{fig:Sc_phonon}
\end{figure}

Figure \ref{fig:Sc_phonon} illustrates that the volume expansion induced by correlation effects captured with DFT+U induces more structural instabilities in Fm$\overline{3}$m Sc$_{0.75}$Ca$_{0.25}$H$_{2.875}$N$_{0.125}$ at 20 and 30 GPa. Since the lattice constants are quite close between the 20 GPa (PBE) and 30 GPa (PBE+U) cases, they can be used to compare the effects of Hubbard parameters on the same lattice parameter as well. In this case, DFT+U appears to somewhat harden the highest energy tetrahedral hydrogen modes and shift the squared frequencies of the lowest frequency modes higher. 

\end{document}